\documentclass[conference]{IEEEtran}
\IEEEoverridecommandlockouts
\usepackage{cite}
\usepackage{amsmath,amssymb,amsfonts}
\usepackage{algorithmic}
\usepackage[dvipdfm]{graphicx}
\usepackage{textcomp}
\usepackage{xcolor}
\usepackage{url}
\usepackage{balance}
\def\BibTeX{{\rm B\kern-.05em{\sc i\kern-.025em b}\kern-.08em
    T\kern-.1667em\lower.7ex\hbox{E}\kern-.125emX}}
\begin{document}
\noindent This paper is the preprint version of the paper ``Classifying DNS Servers based on Response Message Matrix using Machine Learning'' published by IEEE with DOI: 10.1109/CSCI49370.2019.00291.\\

\noindent\copyright{} 2019 IEEE. Personal use of this material is permitted. Permission from IEEE must be obtained for all other uses, in any current or future media, including reprinting/republishing this material for advertising or promotional purposes, creating new collective works, for resale or redistribution to servers or lists, or reuse of any copyrighted component of this work in other works.\\
\newpage

\title{Classifying DNS Servers based on Response Message Matrix using Machine Learning \\
{\normalsize{CSCI-ISCW / Poster Paper}}
}

\makeatletter
\newcommand{\linebreakand}{%
  \end{@IEEEauthorhalign}
  \hfill\mbox{}\par
  \mbox{}\hfill\begin{@IEEEauthorhalign}
}
\makeatother

\author{
\IEEEauthorblockN{1\textsuperscript{st} Keiichi Shima}
\IEEEauthorblockA{\textit{IIJ Innovation Institute, Inc.} \\
Tokyo, Japan \\
keiichi@iijlab.net}
\and
\IEEEauthorblockN{2\textsuperscript{nd} Ryo Nakamura}
\IEEEauthorblockA{\textit{The University of Tokyo} \\
Tokyo, Japan \\
upa@nc.u-tokyo.ac.jp}
\and
\IEEEauthorblockN{3\textsuperscript{rd} Kazuya Okada}
\IEEEauthorblockA{\textit{The University of Tokyo} \\
Tokyo, Japan \\
okada@ecc.u-tokyo.ac.jp}
\linebreakand
\IEEEauthorblockN{4\textsuperscript{th} Tomohiro Ishihara}
\IEEEauthorblockA{\textit{The University of Tokyo} \\
Tokyo, Japan \\
sho@c.u-tokyo.ac.jp}
\and
\IEEEauthorblockN{5\textsuperscript{th} Daisuke Miyamoto}
\IEEEauthorblockA{\textit{The University of Tokyo} \\
Tokyo, Japan \\
daisu-mi@nc.u-tokyo.ac.jp}
\and
\IEEEauthorblockN{6\textsuperscript{th} Yuji Sekiya}
\IEEEauthorblockA{\textit{The University of Tokyo} \\
Tokyo, Japan \\
sekiya@nc.u-tokyo.ac.jp}
}

\maketitle

\begin{abstract}
Improperly configured domain name system (DNS) servers are sometimes used as packet reflectors as part of a DoS or DDoS attack.  Detecting packets created as a result of this activity is logically possible by monitoring the DNS request and response traffic.  Any response that does not have a corresponding request can be considered a reflected message; checking and tracking every DNS packet, however, is a non-trivial operation.  In this paper, we propose a detection mechanism for DNS servers used as reflectors by using a \textit{DNS server feature matrix} built from a small number of packets and a machine learning algorithm.  The F1 score of bad DNS server detection was more than 0.9 when the test and training data are generated within the same day, and more than 0.7 for the data not used for the training and testing phase of the same day.
\end{abstract}


\section{Introduction}
Domain name system (DNS) is one of the most important technologies of the Internet. We can convert a domain name into an IP address using DNS. Without this service, the Internet would not be deployed as widely as it is now. DNS messages are normally built on top of UDP packets. Unlike in TCP, it is easy to forge the source address of UDP packets. As a result, DNS requests with a fake source address can easily be sent to a DNS server. In theory, any DNS server can answer any domain name resolution request; there are no protocol requirements that limit or filter request messages from client nodes. When DNS was invented, malicious activity utilizing DNS servers as packet reflectors was not extensive; however, as the Internet grew, attackers started to use this open operating policy to send traffic to victim nodes by forging DNS message source addresses. To prevent this activity, recent DNS servers have been configured to answer requests originating only from specific client nodes, typically filtered by source IP address. Unfortunately, there are more than a few improperly configured DNS servers in the wild; these are called open resolvers\footnote{DNS Scanning Project: \url{https://dnsscan.shadowserver.org}}. The DNS protocol is still one of the major methods for attacking \cite{durumeric2014-scanning}\cite{alieyan2016}\cite{Kuhrer2015}. In this paper, we propose a method of classifying a DNS server, according to whether or not it is used as a reflector, by monitoring the incoming DNS messages. We collect a series of DNS packets sent from a DNS server and build a feature matrix of the server, assuming that a reflector may have a different packet sequence pattern than that found with a normal DNS server. The preliminary result shows that our method can classify reflectors with an F1 score greater than 0.9 when the test and training data are generated within the same day. The trained model can also classify the data not used for the training and testing phase of the same day with more than 0.7 F1 score.

\section{DNS Server Feature Matrix}

The basic idea behind this proposal originates from \cite{nakamura2018-synpic}. \cite{nakamura2018-synpic} was invented to detect malicious nodes by investigating a series of TCP SYN packets sent from these nodes.  TCP SYN packets are collected based on the source IP addresses of the TCP streams and a feature matrix as an image is generated. In the aforementioned study, it was assumed that the images have different shapes that are dependent on the activities of a malicious host, for example, scanning or DoS. The images generated from SYN packets were used as training data of a deep learning network using a CNN algorithm.

We follow a similar process in our proposal. The difference is that we use DNS response packets received from servers as an input for building the feature matrix.

To apply our method, we first create training data. To split the DNS messages into good messages and suspicious messages, we used the mechanism proposed in \cite{kambourakis2007-detect-dns-amp}. We monitor DNS messages at the boundary of an organization’s network and check all request and response messages. If there is a DNS server being used as a reflector, and it is sending unintended response messages, we will not see any matching request messages sent from within the organization.

The values we used to generate a feature matrix are shown in TABLE~\ref{tab:dns-values}.
\begin{table}[htbp]
\caption{Values of DNS message used to build a feature matrix}
\begin{center}
\begin{tabular}{|l|l|}
\hline
\textbf{Type} & \textbf{Description} \\
\hline
Timestamp & Timestamp of a packet \\
\hline
Port \# & Source port \# of a packet \\
\hline
Size & Size of a DNS message \\
\hline
OPCODE field & Indicating the DNS message type \\
\hline
AA field & Indicating Authoritative Answer or not\\
\hline
TC field & Indicating if a packet is truncated\\
\hline
RD field & Indicating if recursive query is desired\\
\hline
RA field & Indicating if recursive query is available\\
\hline
Z field & Reserved field and should be 0\\
\hline
RCODE field & Indicating result code\\
\hline
QDCOUNT & \# of query items\\
\hline
ARCOUNT & \# of answer records\\
\hline
NSCOUNT & \# of name servers information\\
\hline
AACOUNT & \# of additional records\\
\hline
\end{tabular}
\label{tab:dns-values}
\end{center}
\end{table}

The captured messages are grouped by source IP address (in this case the DNS server IP address), sorted by timestamp, and divided into groups of 100 packets.  Fig.~\ref{fig:dns-server-feature-matrix} shows an example of a DNS server feature matrix.
\begin{figure}[htbp]
\centerline{\includegraphics[width=\columnwidth]{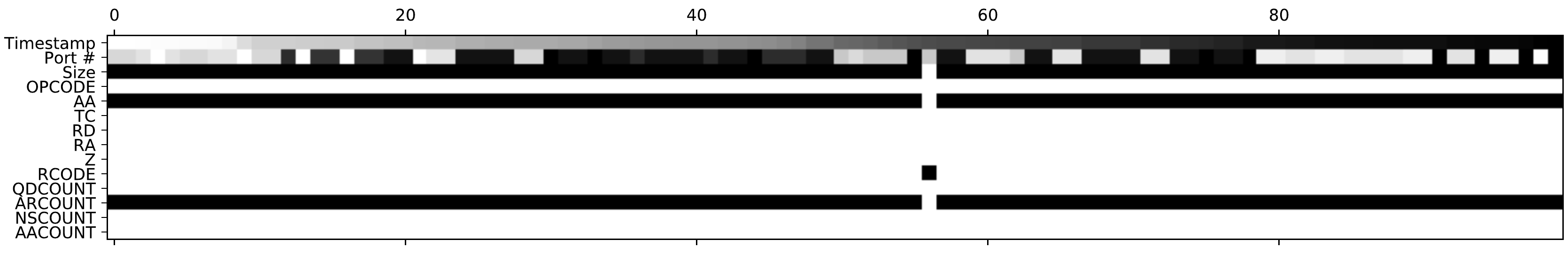}}
\caption{Example of a visualized DNS server feature matrix}
\label{fig:dns-server-feature-matrix}
\end{figure}

The order of rows is the same as the order presented in Table~\ref{tab:dns-values}. The values are normalized per row. Each column indicates one DNS response message. Because  feature matrix is created every 100 packets, the size of columns is 100.

\section{Learning with SVM}

The feature matrix image shown in Fig.~\ref{fig:dns-server-feature-matrix} is based on messages sent from a suspicious DNS server. This particular server kept sending unsolicited DNS response messages; we can guess the behavior by observing the image.  A smoothly changing timestamp row means that messages are being sent periodically. Most packets have the same shape except for source port number. Rows that are almost white or black signify, in most cases, the same values.

Fig.~\ref{fig:good-dns-server-feature-matrix} shows a feature matrix of a good DNS server.
\begin{figure}[htbp]
\centerline{\includegraphics[width=\columnwidth]{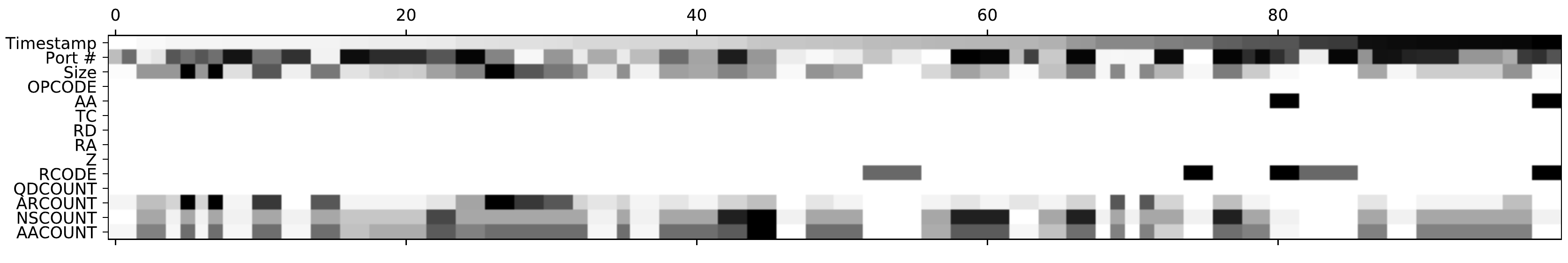}}
\caption{Example of a feature matrix of a good DNS server}
\label{fig:good-dns-server-feature-matrix}
\end{figure}
Different from the case shown in Fig.~\ref{fig:dns-server-feature-matrix}, the fields indicating the number of resource records (such as ARCOUNT) in each response packet have several different values. This is plausible because the contents of DNS request messages sent to a specific DNS server vary according to client; responses may also vary, depending on the request messages.

The datasets used with SVM are a single day data of a certain research network captured between 24\textsuperscript{th} August 2019 and 25\textsuperscript{th} August 2019. The sizes of the datasets are listed in TABLE~\ref{tab:dataset-description}.

\begin{table}[htbp]
\caption{Packet and Matrix counts of datasets}
\begin{center}
\begin{tabular}{|l|l|l|}
\hline
\textbf{Date} & \textbf{\# of Good / Bad DNS pkts} & \textbf{\# of Good / Bad matrices}\\
\hline
24\textsuperscript{th} Aug. & 33,824,531 / 2,863,321 & 323,269 / 28,291 \\
\hline
25\textsuperscript{th} Aug. & 30,238,481 / 1,148,935 & 291,730 / 6,105 \\
\hline
\end{tabular}
\end{center}
\label{tab:dataset-description}
\end{table}

The selected hyper-parameters were penalty = 10, gamma = 0.01, and kernel = rbf, using grid search.  The model was trained and tested with 20,000 randomly selected good matrices, and 80\% of bad matrices for each day. For example, when using the dataset of 24\textsuperscript{th}, we randomly sampled 20,000 matrices from good matrices and $0.8 \times 28,291 = 22,632$ bad matrices. The ratio of training data and test data was 0.8 and 0.2.

TABLE~\ref{tab:res-0824-samples} and \ref{tab:res-0825-samples} present the classification results of sampled data for each day. As we can see from the tables, as long as we focus on the sampled data, the classification accuracy is high enough.

\begin{table}[htbp]
\caption{Classification result of sample data of 24\textsuperscript{th} Aug.}
\begin{center}
\begin{tabular}{l|r|r|r|r}
& Precision & Recall & F1-score & Support \\
\hline
Good          & 1.00 & 1.00 & 1.00 & 3,987 \\
Bad           & 1.00 & 1.00 & 1.00 & 4,540 \\
\hline
Accuracy      &      &      & 1.00 & 8,527 \\
Macro Avg.    & 1.00 & 1.00 & 1.00 & 8,527 \\
Weighted Avg. & 1.00 & 1.00 & 1.00 & 8,527 \\
\end{tabular}
\end{center}
\label{tab:res-0824-samples}
\end{table}

\begin{table}[htbp]
\caption{Classification result of sample data of 25\textsuperscript{th} Aug.}
\begin{center}
\begin{tabular}{l|r|r|r|r}
& Precision & Recall & F1-score & Support \\
\hline
Good          & 1.00 & 1.00 & 1.00 & 3,993 \\
Bad           & 0.98 & 1.00 & 0.99 &   984 \\
\hline
Accuracy      &      &      & 1.00 & 4,977 \\
Macro Avg.    & 0.99 & 1.00 & 0.99 & 4,977 \\
Weighted Avg. & 1.00 & 1.00 & 1.00 & 4,977 \\
\end{tabular}
\end{center}
\label{tab:res-0825-samples}
\end{table}

Next, we evaluated the rest of the data in the datasets not used in the training phase for each day. The results are shown in TABLE~\ref{tab:res-0824-rest} and \ref{tab:res-0825-rest}.

\begin{table}[htbp]
\caption{Classification result of unused data of 24\textsuperscript{th} Aug.}
\begin{center}
\begin{tabular}{l|r|r|r|r}
& Precision & Recall & F1-score & Support \\
\hline
Good          & 1.00 & 1.00 & 1.00 & 303,269 \\
Bad           & 0.85 & 1.00 & 0.92 &   5,659 \\
\hline
Accuracy      &      &      & 1.00 & 308,928 \\
Macro Avg.    & 0.92 & 1.00 & 0.96 & 308,928 \\
Weighted Avg. & 1.00 & 1.00 & 1.00 & 308,928 \\
\end{tabular}
\end{center}
\label{tab:res-0824-rest}
\end{table}

\begin{table}[htbp]
\caption{Classification result of unused data of 25\textsuperscript{th} Aug.}
\begin{center}
\begin{tabular}{l|r|r|r|r}
& Precision & Recall & F1-score & Support \\
\hline
Good          & 1.00 & 1.00 & 1.00 & 271,730 \\
Bad           & 0.54 & 1.00 & 0.70 &   1,221 \\
\hline
Accuracy      &      &      & 1.00 & 272,951 \\
Macro Avg.    & 0.77 & 1.00 & 0.85 & 272,951 \\
Weighted Avg. & 1.00 & 1.00 & 1.00 & 272,951 \\
\end{tabular}
\end{center}
\label{tab:res-0825-rest}
\end{table}

\begin{figure}[htp]
\centerline{\includegraphics[width=\columnwidth]{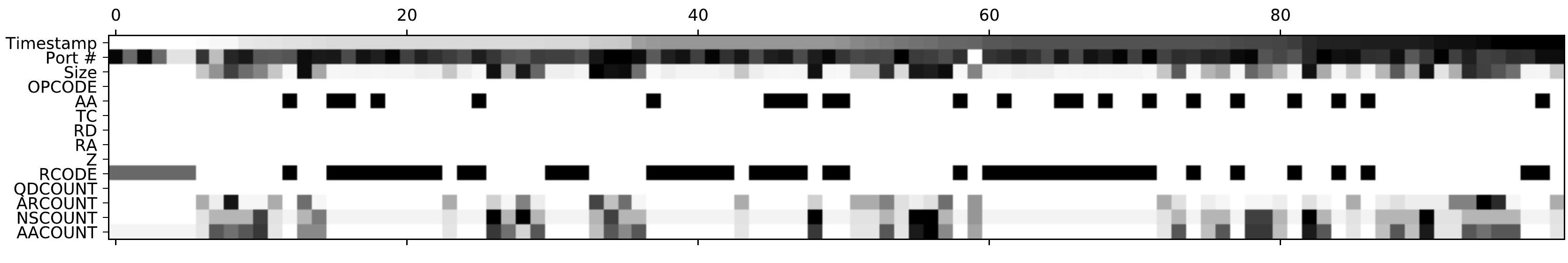}}
\caption{Example of a feature matrix of an uncertain server}
\label{fig:uncertain-feature-matrix}
\end{figure}

The precision values are decreased on both days that means more false positive results are seen.  The F1-score on 24\textsuperscript{th} is still acceptable, however, the score on 25\textsuperscript{th} is largely degraded. Since the recall values of both good matrices and bad matrices are kept high, we can still detect bad matrices with high enough probability. 

Fig~\ref{fig:uncertain-feature-matrix} shows a feature matrix of a DNS server labeled as a bad matrix which we didn't see any request messages for the response messages sent from the DNS server. The shape looks quite similar to that of a good feature matrix shown in Fig~\ref{fig:good-dns-server-feature-matrix} in the sense that the contents of the response messages have a wide variety of patterns. The server shown in Fig~\ref{fig:uncertain-feature-matrix} was one of the DNS servers of the host organization of the datasets where we captured the packets. Considering the quality of the security operators of the organization, it is unlikely that the server was used as a reflector.  Our guess is that the request messages went to the server using the different path where we were monitoring the traffic.

Cleansing of source data when using machine learning techniques is one of the important phases to achieve reliable results, and at the same time, it is one of the hardest tasks, especially the size of the data is big and the contents are dynamic and changing. Since the Internet is open system and the traffic trends are undoubtedly changing every day, assigning correct labels to training dataset is not an easy task. In this preliminary experiments, we did not perform intensive data cleansing because of lack of time.  For example the matrix pattern shown in Fig~\ref{fig:uncertain-feature-matrix} may be a benign pattern. We continue to investigate the contents of the dataset in more detail to achieve better labels.

\section{Conclusion}
\balance

We attempted to classify DNS servers according to whether or not they were being used as reflectors by capturing a small number of DNS response messages sent from them.  We used a method similar to the one proposed in \cite{nakamura2018-synpic} to build a DNS server feature matrix.  The preliminary results of classification using SVM show sufficient precision as long as training and test data from the same day is used.  At this moment, the trained model does not show as high classification result when applied to the rest of the data which are not used for training and testing. One possible reason is the improper labeling of the data.  As we described, we labeled each matrix based on the technique described in \cite{kambourakis2007-detect-dns-amp}. The method can find all the unsolicited DNS response messages assuming we can monitor the entire DNS message exchanges. In our preliminary experiments, we were seeing unsolicited DNS response messages sent from the servers located inside the host organization, which may be benign servers. Assigning correct labels to data is important when using the data as a training dataset for machine learning algorithms.  We plan to investigate the contents in more detail to create better training datasets.

The classification method we used in this paper was SVM.  SVM is a simple and easy-to-use tool for data analysis, however, we recently have more advanced algorithms. Therefore, in the future, we plan on making the results more stable by investigating data and matrix generation approaches (e.g. what values to use to build a matrix) and also by investigating classification algorithms (including deep learning technologies) to achieve superior performance.

\section*{Acknowledgement}

This work was supported by JST CREST Grant Number JPMJCR1783, Japan.

\bibliographystyle{./bibliography/IEEEtran}
\bibliography{./bibliography/bibliography,./bibliography/IEEEabrv,./bibliography/IEEEexample}

\end{document}